# Flat AgTe Honeycomb Monolayer with Topologically Nontrivial States

Bing Liu,[†,‡,§] Jian Liu,[†,‡,§] Guangyao Miao,[†,‡] Siwei Xue,[†,‡] Shuyuan Zhang,[†,‡] Lixia Liu,[†,‡] Xiaochun Huang,[†,‡] Xuetao Zhu,[†,‡] Sheng Meng,[†,‡] Jiandong Guo,[†,‡,//] Miao Liu,[*,†,⊥,//] and Weihua Wang[*,†]

[†]Beijing National Laboratory for Condensed Matter Physics and Institute of Physics, Chinese Academy of Sciences, Beijing 100190, China

[‡]School of Physical Sciences, University of Chinese Academy of Sciences, Beijing 100190, China

[//]Songshan Lake Material Laboratory, Dongguan, Guangdong 523808, China

[⊥]Center of Materials Science and Optoelectronics Engineering，University of Chinese Academy of Sciences, Beijing 100049, China

[§]These authors contributed equally to this work.

*Email: mliu@iphy.ac.cn; weihuawang@iphy.ac.cn

The intriguing properties, especially Dirac physics in graphene, have inspired the pursuit of two-dimensional materials in honeycomb structure. Here we achieved a monolayer transition metal monochalcogenide AgTe on Ag(111) by tellurization of the substrate. High-resolution scanning tunneling microscopy, combined with low-energy electron diffraction, angle-resolved photoemission spectroscopy, and density functional theory calculations, demonstrates the planar honeycomb structure of AgTe. The first principle calculations further reveal that, protected by the in-plane mirror reflection symmetry, two Dirac node-line Fermions exist in the electronic structures of free-standing AgTe when spin-orbit coupling (SOC) is ignored. While in fact the SOC leads to the gap opening, and resulting in the emergence of the topologically nontrivial quantum spin Hall edge state. Importantly, our experiments evidence the chemical stability of the monolayer AgTe in ambient conditions. It is possible to study AgTe by more *ex-situ* measurements and even to apply it in novel electronic devices.

The discovery of graphene and its intriguing properties have inspired the pursuit of two-dimensional (2D) materials in honeycomb structure.[1-4] It has been proposed that electrons hopping in a honeycomb structure can realize quantum anomalous Hall effect.[5] Various 2D Xenes composed of monoelemental atoms (IVA and VA families in the periodic table) arranged in a honeycomb structure were predicted theoretically or investigated experimentally to be quantum spin Hall materials protected by time reversal symmetry, such as silicene, germanene, stanine, antimonene, and bismuthene.[6-12] In addition to monoelemental 2D honeycomb materials, some 2D compounds with honeycomb-like structure possess exotic electronic structures.[8, 13-16] For example, bilayers of perovskite-type transition-metal oxides grown along the [111] crystallographic axis exhibit buckled honeycomb structure, and have been predicted to show topologically nontrivial properties.[17, 18]

As a large family of 2D compound materials, the layered transition metal chalcogenides (TMCs) exhibit emergent complex properties, such as enhanced optoelectronic properties, topological properties and superconductivity.[19-32] Almost all TMCs that have been explored experimentally are dichalcogenides: the monolayer structure contains a layer of transition metal sandwiched by two layers of chalcogen atoms.[25, 33, 34] It is particularly attractive to explore the TMCs with planar honeycomb structures, since the symmetry, including crystal symmetry and time reversal symmetry, may lead to topologically nontrivial electronic states of the materials,[27, 28, 35, 36] *e.g.*, monolayer CuSe in planar honeycomb structure have been predicted to possess Dirac node-line Fermions (DNLFs) protected by mirror reflection symmetry.[37] Recently, the monolayer CuSe with the honeycomb structure, but buckled, has been realized in experiments.[37, 38] It is still challenging to experimentally obtain TMCs with planar honeycomb structure.

Here by direct tellurization of Ag(111), we successfully synthesized high-quality flat AgTe monolayer in honeycomb structure on Ag(111) substrate, which is further investigated by scanning tunneling microscopy (STM), low-energy electron diffraction (LEED), angle-resolved photoemission spectroscopy (ARPES) and density functional theory (DFT) calculations. The X-ray photoelectron spectroscopy (XPS) analyses indicate that chemical bonds exist between Te and Ag atoms at the epitaxial interface. Importantly, the stability of the AgTe layer in ambient conditions is evidenced. More importantly, our DFT calculations reveal that free-standing monolayer AgTe has two DNLFs protected by mirror reflection symmetry when spin-orbit coupling (SOC) is ignored. When SOC is considered, gaps are opened around the Dirac nodal lines, and topologically nontrivial quantum spin Hall edge state emerges.

When a small amount of Te atoms [≤ 0.083 monolayer (ML), here 1 ML refers to the atomic density of Ag (111) surface] are deposited on Ag(111) substrate held at room temperature, a disordered phase is observed in STM, in which Te atoms adsorb in a random way. As Te coverage increasing (~0.125 ML), an ordered phase appears on the surface and its area increases with Te coverage. The ordered phase is 0.15 nm higher than the disordered phase on the same Ag(111) terrace. (See Supporting Information Figure S1 for the whole preparation process.) At the Te coverage of ~0.33 ML, all the Ag(111) terraces are covered by the ordered phase, and a typical large-scale STM image is shown in Figure 1a. The height difference between the nearest terraces is 0.24 nm, corresponding to the intrinsic height of a single Ag(111) step, indicating the epitaxial ordered phase on Ag terraces is a single atomic layer. LEED measurements show that in addition to the diffractive patterns of Ag(111) substrate, the ordered phase has extra $(\sqrt{3} \times \sqrt{3})$ spots along the $\langle 11\bar{2} \rangle$ directions (Figure 1b). The large-scale STM image and sharp LEED patterns also demonstrate the high quality of the ordered phase.

Figure 1c shows an atom-resolved STM image of the ordered phase. In accordance with the LEED results, atoms in the ordered phase are arranged in a periodic superstructure with 3-fold symmetry. As shown in the upper panel of Figure 1d, the superstructure has a periodicity of 5.0 Å alone the $\langle 11\bar{2} \rangle$ directions, which is $\sqrt{3}$ times of the Ag(111) lattice constant. [The lattice constant of Ag(111) is 2.88 Å.]

Detailed investigation of Figure 1c further indicates that there are two types of atoms with different apparent height in STM image (blue arrows in Figure 1c). In the lower panel of Figure 1d, the difference of the apparent height of two nearest atoms is measured to be 6.2 pm at -0.03 V. Note that such apparent height difference is determined by electronic states rather than geometrical height. The higher atoms appear as bright triangles sitting in a triangular sublattice, while the lower ones appear as dark triangles. The apexes of the bright and dark triangles are connected to each other, forming a honeycomb structure (the orange hexagon in Figure 1c). A unit cell (the cyan rhombus in Figure 1c) contains two atoms with different apparent height, corresponding to a nominal atomic coverage of 2/3 ML. Since only 1/3 ML Te atoms were deposited on the substrate, we naturally suggest another 1/3 ML atoms are Ag. Therefore, Te and Ag atoms form the honeycomb structure in a 1:1 ratio, *i.e.*, AgTe compound.

The epitaxial AgTe on Ag(111) were further studied by XPS. From the preparation chamber to XPS chamber, the sample is exposed to atmosphere. We have verified by STM experiments that a moderate annealing in ultra-high vacuum (UHV) conditions at 500 K for 8 min will remove all the adsorbates on AgTe surface, and the atomic structure of AgTe are not affected (discussed below). Figure 1e and 1f show the 3$d$ core level of Te and Ag in monolayer AgTe on Ag(111) that has been transferred to XPS chamber and further annealed at 500 K for 10 min in UHV conditions. Obviously, compared to elemental Te whose 3$d_{3/2}$ peak is located at the binding energy (BE) of 583.3 eV, the Te 3$d_{3/2}$ peak in AgTe is shifted to 582.8 eV, indicating Ag-Te bonds are formed in the AgTe sample. However, the BE of Ag 3$d_{3/2}$ and 3$d_{5/2}$ peaks are identical to elemental Ag, locating at 374.3 eV and 368.3 eV, respectively. This is explained by the fact that XPS collects signals from several surface layers. Therefore, the Ag signal detected by XPS is dominated by Ag(111) substrate.

To further elucidate the experimental results above, we carried out the first-principles calculations on AgTe adsorbed on Ag (111) substrate. By comparing the total energy of different adsorption models with 3-fold symmetry, in which the Te and Ag atoms in AgTe are initially placed at high-symmetry sites, such as top,

*fcc*-hollow, and *hcp*-hollow sites, the model with all atoms at *hcp* sites is found to be the most stable one. The top and side views of the optimized adsorption model are shown in Figure 2a and 2b. (See Supporting Information Figure S2 for other models.)

The optimized adsorption model takes a honeycomb structure with Ag and Te atoms arranged alternately, forming a $(\sqrt{3}\times\sqrt{3})$ supercells on Ag(111) substrate. The lattice constant is 5.0 Å, in agreement with STM and LEED results. Moreover, according to the optimized adsorption model, Te and Ag atoms in AgTe have the same height, *i.e.*, the honeycomb AgTe is perfectly planar, in contrast to the buckled CuSe.[37] Figure 2c shows the simulated STM image (top panel) by DFT calculations, completely coincident with the STM image in experiment (bottom panel), where the higher and lower atoms in the simulated image are Ag and Te atoms, respectively.

To illustrate the electronic structure of AgTe in experiment, we conducted ARPES characterization of AgTe on Ag(111) substrate. Figure 2d indicates that the surface states of Ag(111) is totally modulated by the epitaxy of monolayer AgTe.[39] In addition to the bands with strong intensity from -6 eV to -4 eV which are donated by Ag(111) substrate, near the Fermi level, electron-like bands of Ag(111) surface disappears, and hole-like bands at energy above -2 eV are detected instead. Our DFT calculations of monolayer AgTe on Ag(111) with SOC effect demonstrate that there are two hole-like bands contributed by AgTe (Figure 2e). Limited by the resolution of our ARPES system, these two hole-like bands cannot be obviously distinguished. The consistency of ARPES and DFT calculations results further confirms that the prepared sample is monolayer AgTe on Ag(111) substrate.

Environmental stability is important for the practical applications of 2D materials in electronic devices, and it is also the prerequisite for a material to be transferred and measured by *ex-situ* techniques without a coating layer. To study the stability of AgTe in ambient conditions, we exposed the epitaxial AgTe to the atmosphere at room temperature for 30 minutes. Then the sample was reloaded into the UHV chamber, and annealed at 500 K for 8 minutes to remove possible adsorbates. Afterwards, it was measured by STM again, and the typical images are shown in Figure 3a and 3b.

Figure 3a reveals the AgTe terraces are quite similar to the as-grown sample (Figure 1a), except a minimal amount impurity near the step. Atom-resolved STM image in Figure 3b proves the crystal structure of AgTe is unchanged. This experiment evidences that the exposure to atmosphere does not affect AgTe film chemically.

We also compared the XPS spectra of the $3d$ core level of Te and Ag on as-transferred and annealed AgTe on Ag(111), as well as bulk Te (polycrystal, source material used in this experiment) and bulk Ag (polycrystal) to reveal the possible reactions of monolayer AgTe in ambient conditions (Figure S3). The Te $3d_{3/2}$ peak in both as-transferred and annealed samples is at the same BE of 582.8 eV, indicating AgTe does not react with oxygen or water severely in ambient conditions. We also detected weak O signal at BE of 530.7 eV on the as-transferred sample, but the O signal was thoroughly eliminated after annealing, indicating that oxygen atom or molecules are weakly adsorbed on AgTe, and moderate annealing can clean the AgTe layer totally. Above all, AgTe is fairly stable in the air under room temperature, enabling other *ex-situ* measurements on this material.

The AgTe monolayer grown on Ag(111) has a fairly flat honeycomb structure, and, hence, services as an ideal material system to investigate Dirac fermions that are protected by highly symmetric honeycomb configuration, such as the mirror reflection symmetry with respect to the $M_{xy}$ plane. To avoid the influence of Ag(111) substrate and to isolate the electronic states of AgTe, we simplify this system to a free-standing monolayer AgTe model. We have verified the free-standing AgTe prefers the planar structure rather than the buckled one by fully relaxing the atoms. Furthermore, the absence of imaginary phonon modes is a significant indication that free-standing AgTe in the planer structure is thermodynamically stable (Figure S4). Figure 4a shows the calculated band structure of the free-standing AgTe without considering SOC. Near the Fermi level, there are two hole-like bands marked as $\alpha$ and $\beta$ bands, and one electron-like band denoted as $\gamma$ band. The $\gamma$ band goes across $\alpha$ and $\beta$ bands linearly at approximately -0.22 eV and -0.46 eV without opening gaps. It is noted that all bands are double degenerate, hence the gapless nodes correspond to the fourfold Dirac fermions. The momentum distributions of both two Dirac fermions are two Dirac nodal lines (NL1 and NL2 in Figure 4b).

We further illustrated whether both two DNLFs in monolayer AgTe are protected by mirror reflection symmetry. From the projected band structures, we judge that $\alpha/\beta$ bands are mainly contributed by the in-plane orbitals (Ag $d_{xy}/d_{x^2-y^2}$, Te $p_x/p_y$), and $\gamma$ band is attributed to the out-of-plane orbital (Te $p_z$) (Figure S5). Along the Γ-K direction, the little group of AgTe is $C_s$ with mirror reflection symmetry with respect to the $M_{xy}$ plane. And along the Γ-M direction, the little group of AgTe is $C_{2v}$ with mirror reflection symmetry with respect to the $M_{xy}$ and $M_{yz}$ planes. Like CuSe and Cu$_2$Si which have similar lattice and energy band structures, the $M_{xy}$ mirror parities of the in-plane $\alpha$ and $\beta$ bands are even ("+"), while that of the out-of-plane $\gamma$ band is odd ("-").[37, 40] The cross of bands with opposite parities indicates both two DNLFs are protected by mirror reflection symmetry with respect to the $M_{xy}$ plane.

The mirror reflection symmetry $M_{xy}$ will be broken when considering SOC effect which leads to the coupling of in-plane and out-of-plane orbitals, as a result, energy gaps are opened at Dirac points, ensuring us to investigate the topological properties of the DNLFs. Figure 4c shows the band structure of free-standing AgTe when considering SOC, exhibiting energy gaps along the M-Γ-K direction which are 0.35 eV, 0.27 eV, 0.23 eV, and 0.24 eV, respectively. So large gaps are beneficial to accommodate edge states. Figure 4d shows the edge states obtained by projecting bulk states on the zigzag edge (Γ-K direction). For NL1, obviously, a pair of helical edge states appear in the gap (circled by the green dashed line in Figure 4d), signaling quantum spin Hall state.[27, 28, 36, 41] The topological invariant $\mathcal{Z}_2$ is calculated to be one (Figure S8a), confirming the topologically nontrivial property of the edge states. For NL2, there is an edge band connecting the bulk bands on two sides of the energy gap and reaches edges of the Brillouin zone (circled by the purple dashed line in Figure 4d). The calculated topological invariant $\mathcal{Z}_2$ is zero, similar to the case in compressed black phosphorus.[42] Although this edge state is topologically trivial, it can be stable in impurity-free system.

On the other hand, the buckling of AgTe plane can also break the mirror reflection symmetry $M_{xy}$, and induce energy gaps, as shown in Figure S6. Moreover, we found one of the two Dirac cones along the Γ-M direction keeps gapless, because of the protection of mirror reflection symmetry $M_{yz}$ (discussed in Supporting Information). This result further confirms the mirror-symmetry-protected properties of these two DNLFs.

Due to the relatively strong interaction between AgTe and Ag(111) substrate in the out-of-plane direction, $\gamma$ band that is contributed by $p_z$ orbitals of Te atoms disappears in real system (Figure 2d and 2e). Therefore, the topological edge states disappear under the influence of Ag(111) substrate. We suggest to intercalate other atoms or molecules to decouple AgTe from Ag(111) substrate, or to seek other substrates that have weak interaction with AgTe, to obtain intrinsic electronic structures of monolayer AgTe and investigate its topological edge states in future works.

In summary, we have successfully realized flat AgTe monolayer in honeycomb structure on Ag (111) substrate. DFT calculations propose that the free-standing monolayer AgTe may have hosted two DNLFs protected by its mirror reflection symmetry, while the SOC effect leads to the opening of the energy gaps around both two Dirac nodal lines, and topologically nontrivial edge states appear in one of the gaps, with the other gap possessing the trivial edge states. We have also verified experimentally the chemical stability of monolayer AgTe in ambient conditions. Further investigations are expected for the monolayer AgTe as a promising system of quantum spin Hall effect for future applications of novel electronics devices.

This work was supported by the National Key Research & Development Program of China (Nos. 2016YFA0300600, 2017YFA0303600, and 2016YFA0202300), and the Hundred Talents Program of the Chinese Academy of Sciences. X.Z. was partially supported by the Youth Innovation Promotion Association of Chinese Academy of Sciences.

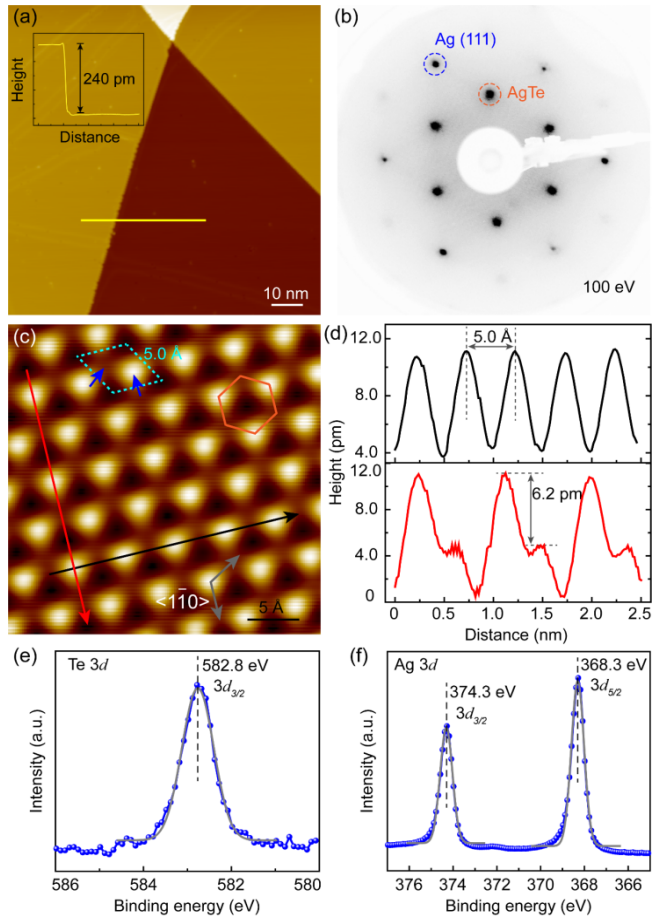

**Figure 1.** (a) Large scale STM image of monolayer AgTe on Ag(111) substrate (-1.0 V, 0.05 nA). All the terraces shown here are covered by AgTe. Inset: The height profile along the yellow line. (b) LEED pattern of monolayer AgTe on Ag(111) substrate. Blue circle marks the diffractive pattern of Ag(111) substrate, and orange circle marks the ($\sqrt{3}\times\sqrt{3}$)R30° diffractive pattern of AgTe. (c) Zoom-in STM image of monolayer AgTe on Ag(111) (-0.03 V, 1.0 nA). The honeycomb structure is highlighted by the orange hexagon, and the blue arrows indicate the two atoms with different apparent heights within a unit cell, marked by the cyan rhombus. Grey arrows indicate the <1$\bar{1}$0> orientations of Ag(111) substrate. (d) Height profiles alone the black and red lines in (c). (e, f) Te 3$d$ and Ag 3$d$ core level XPS spectra of AgTe on Ag(111) substrate.

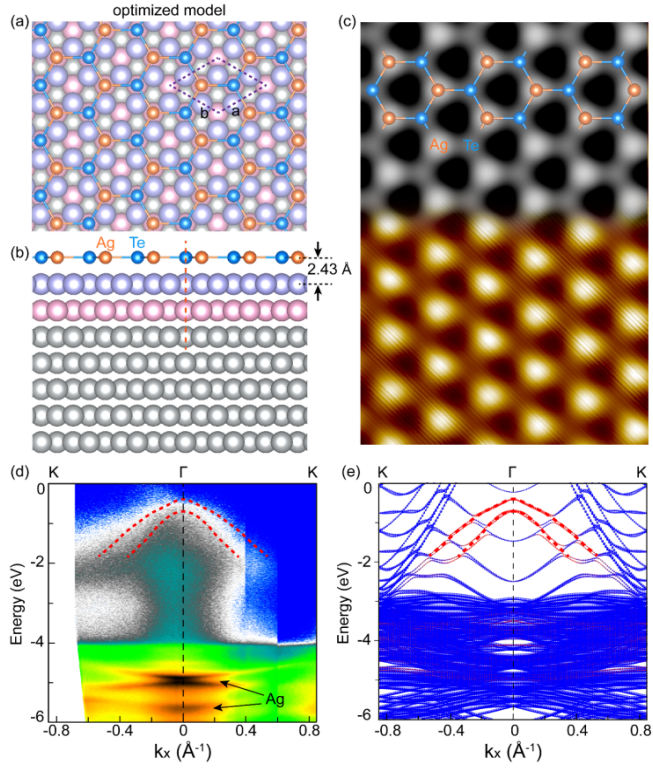

**Figure 2**. Top (a) and side (b) views of the optimized atomic model of monolayer AgTe on Ag (111) substrate, with all Ag and Te atoms at *hcp* sites in a planar honeycomb structure. The dashed rhombus in (a) indicates the unit cell of AgTe. Orange and blue balls represent Ag and Te atoms in AgTe, respectively. Purple and pink balls represent the first and the second layer Ag atoms of Ag(111) substrate, and silvery balls represent the lower Ag atoms of the substrate. (c) Simulated (top panel) and experimental (bottom panel) STM images with atomic model superposed on. (d) ARPES intensity plots along the K-Γ-K direction of monolayer AgTe on Ag(111) substrate. (e) DFT calculated band structure along the K-Γ-K direction of monolayer AgTe on Ag(111) substrate with SOC. Red and blue ingredients are contributed by AgTe and Ag(111) substrate, respectively.

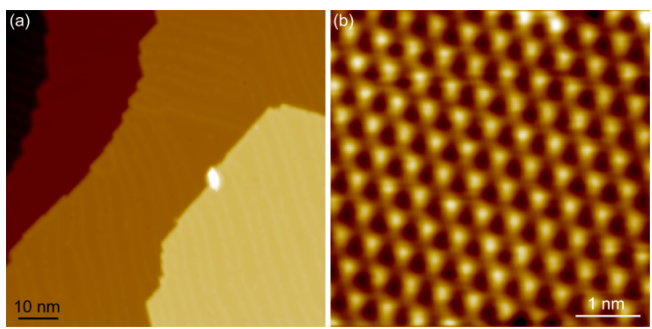

**Figure 3**. Large scale STM image (-1.0 V, 0.05 nA) (a) and zoom-in STM image (-30 mV, 1.0 nA) (b) of the monolayer AgTe on Ag(111) after exposure on atmosphere for 30 minutes and then annealing at 500 K for 8 minutes.

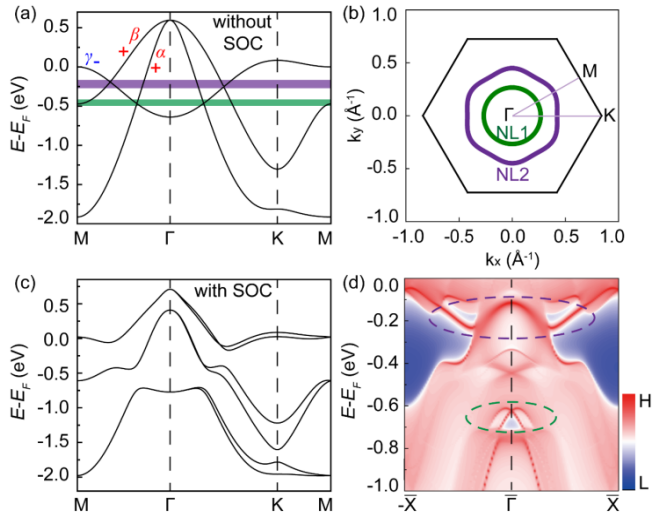

Figure 4. (a) Band structure of free-standing monolayer AgTe without SOC. The parities of $\alpha$, $\beta$ and $\gamma$ bands are shown. (b) Distribution of two Dirac node lines (NL1 and NL2) in momentum space, corresponding to the green and purple shaded area in (a). The black hexagon lines denote the first Brillouin zone. (c) Band structure of free-standing monolayer AgTe with SOC. (d) Edge states of the semi-infinite AgTe along the zigzag edge. The green and purple dashed lines circle the edge states in the gaps opened around NL1 and NL2, respectively.

Methods:

The sample preparation and STM investigations were conducted in an ultrahigh-vacuum molecular beam epitaxy (MBE)-STM joint system (Unisoku) with a base pressure better than $1.0\times10^{-10}$ mbar. The Ag (111) substrate was cleaned by cycles of Ar$^+$ sputtering and annealing, until a clean and uniform surface was obtained. Te (99.999+%, Alfa Aesar) were evaporated from a Knudsen cell at ~500 K, and deposited on Ag(111) substrate held at room temperature. The deposition rate of Te was kept at ~0.083 ML per minute. The STM results were obtained at 77 K and 4.9 K. The XPS measures were performed at an X-ray photoelectron spectrometer of ThermoFisher Scientific ESCALAB 250X. The Al Kα source with the photon energy of 1486.6 eV was used. The BE was calibrated with respect to the pure bulk Au $4f_{7/2}$ (BE=84.0 eV) and Cu $2p_{3/2}$ (BE=932.7 eV) lines. The BE is referenced to the Fermi level ($E_f$) calibrated by using pure bulk Ni as $E_f$=0 eV. It should be noted that the LEED, XPS, and ARPES measurements were *ex situ*. After the exposure to atmosphere during the transfer, the sample was annealed at 500 K for 8 min to 10 min in the LEED/XPS/ARPES UHV chambers. As tested by STM, this procedure could remove all the adsorbates from the sample surface.

The first-principles calculations were performed based on density functional theory as implemented in the Vienna *ab initio* simulation package.[43, 44] The generalized gradient approximation with Perdew-Burke-Ernzerhof functional was employed to describe exchange-correlation potential.[45] The cutoff energy of plane-wave of 500 eV was set in all the calculations. From the first-principles results, one Ag *s* orbital, five Ag *d* orbitals and three Te *p* orbitals were used to construct the maximally localized Wannier functions.[46] The surface states were obtained by using the iteration Green's function method as implemented in Wanniertools package.[47]